\documentclass[onecolumn]{aa}

\usepackage{graphicx}
\usepackage{txfonts}

\newcommand{\Nside}{\ensuremath{N_{\mathrm{side}}}} 
\newcommand{\Npix}{\ensuremath{N_{\mathrm{pix}}}}   
\newcommand{\Ntau}{\ensuremath{N_{\tau}}}   
\newcommand{\vA}{\vec{A}}
\newcommand{\vB}{\vec{B}}
\newcommand{\vM}{\vec{M}}
\newcommand{\vN}{\vec{N}}
\newcommand{\vP}{\vec{P}}
\newcommand{\vS}{\vec{S}}

\newcommand{\vd}{\vec{d}}
\newcommand{\vn}{\vec{n}}
\newcommand{\vs}{\vec{s}}

\newcommand{\vx}{\vec{x}}
\newcommand{\vshat}{\vec{\hat{s}}}
\newcommand{\vxhat}{\vec{\hat{x}}}
\newcommand{\vnu}{\vn_{u}}
\newcommand{\vNu}{\vN_{u}}
\newcommand{\tr}{^{\mathrm{T}}} 
\newcommand{\inv}{{}^{-1}}      
\newcommand{\muK}{\ensuremath{\mathrm{\mu K}}}
\newcommand{\mo}{\ensuremath{^{-1}}}

\begin{document}

\title{Making Sky Maps from Planck Data}

\author{%
  M.\ A.\ J.\ Ashdown\inst{1,2}
  \and
  C.\ Baccigalupi\inst{3,4}
  \and
  A.\ Balbi\inst{5}
  \and
  J.\ G.\ Bartlett\inst{6}
  \and
  J.\ Borrill\inst{7,8}
  \and
  C.\ Cantalupo\inst{8,7}
  \and
  G.\ de Gasperis\inst{5}
  \and
  K.\ M.\ G\'{o}rski\inst{9,10,11}
  \and
  E.\ Hivon\inst{10,14}
  \and
  E.\ Keih\"{a}nen\inst{12,13}
  \and
  H.\ Kurki-Suonio\inst{12}
  \and
  C.\ R.\ Lawrence\inst{9}
  \and
  P.\ Natoli\inst{5}
  \and
  T.\ Poutanen\inst{12,13}
  \and
  S.\ Prunet\inst{14}
  \and
  M.\ Reinecke\inst{15}
  \and
  R.\ Stompor\inst{7,8,6}
  \and
  B.\ Wandelt\inst{16,17}\\
  (The Planck CTP Working Group)
}

\offprints{M. A. J. Ashdown, \email{maja1@mrao.cam.ac.uk}}

\institute{%
  Astrophysics Group, Cavendish Laboratory, J J Thomson Avenue,
  Cambridge CB3 0HE, United Kingdom.
  \and
  Institute of Astronomy, Madingley Road, Cambridge CB3 0HA,
  United Kingdom.
  \and
  Institut f\"{u}r Theoretische Astrophysik,
  Universit\"{a}t Heidelberg, Albert-\"{U}berle-Str. 2,
  D-69120, Heidelberg, Germany.
  \and
  SISSA/ISAS, Via Beirut 4, I-34014 Trieste, and INFN,
  Sezione di Trieste, Via Valerio 2, I-34127, Italy
  \and
  Dipartimento di Fisica, Universit\`{a} di Roma ``Tor Vergata'',
  via della Ricerca Scientifica 1, I-00133 Roma, Italy.
  \and
  Laboratoire Astroparticule \& Cosmologie,
  11 place Marcelin Berthelot, 75231 Paris Cedex 05, France
  (UMR 7164 CNRS, Universit\'e Paris 7, CEA, Observatoire de Paris).
  \and
  Computational Research Division, Lawrence Berkeley National
  Laboratory, Berkeley CA 94720, U.\ S.\ A.
  \and
  Space Sciences Laboratory,
  University of California Berkeley, Berkeley CA 94720, U.\ S.\ A.
  \and
  Jet Propulsion Laboratory, California Institute of Technology, 4800 Oak
  Grove Drive, Pasadena CA 91109, U.\ S.\ A.
  \and
  California Institute of Technology, Pasadena CA 91125, U.\ S.\ A.
  \and
  Warsaw University Observatory, Aleje Ujazdowskie 4, 00478 Warszawa, Poland.
  \and
  University of Helsinki, Department of Physical Sciences,
  P. O. Box 64, FIN-00014 Helsinki, Finland.
  \and
  Helsinki Institute of Physics, P.\ O.\ Box 64, FIN-00014 Helsinki,
  Finland.
  \and
  Institut d'Astrophysique de Paris, 98 bis Boulevard Arago,
  F-75014 Paris, France.
  \and
  Max-Planck-Institut f\"{u}r Astrophysik, Karl-Schwarzschild-Str.~1,
  D-85741 Garching, Germany.
  \and
  Department of Physics, University of Illinois at
  Urbana-Champaign, 1110 West Green Street, Urbana IL 61801, U.\ S.\ A.
  \and
  Department of Astronomy, University of Illinois at
  Urbana-Champaign, 1002 West Green Street, Urbana IL 61801, U.\ S.\ A.
}

\date{Received date / Accepted date}

\abstract
    {}
    {We compare the performance of multiple codes written by different
    groups for making polarized maps from Planck-sized, all-sky cosmic
    microwave background (CMB) data.  Three of the codes are based on
    a destriping algorithm; the other three are implementations of an
    optimal maximum-likelihood algorithm.}
    {Time-ordered data (TOD) were simulated using the Planck Level-S
    simulation pipeline.  Several cases of temperature-only data were
    run to test that the codes could handle large datasets, and to
    explore effects such as the precision of the pointing data.  Based
    on these preliminary results, TOD were generated for a set of four
    217~GHz detectors (the minimum number required to produce I, Q,
    and U maps) under two different scanning strategies, with and
    without noise.}
    {Following correction of various problems revealed by the early
    simulation, all codes were able to handle the large data volume
    that Planck will produce.  Differences in maps produced are small
    but noticeable; differences in computing resources are large.}
    {}

\keywords{Cosmology: cosmic microwave background -- Methods: data
  analysis}
    
\maketitle

\section{Introduction}

Cosmic microwave background (CMB) observations have driven a
remarkable advance in cosmology over the past decade (Smoot et
al.~\cite{smoot02}; de~Bernardis et al.~\cite{debernardis00}; Hanany
et al.~\cite{hanany00}; Beno\^{\i}t et al.~\cite{benoit03}; Bennett et
al.~\cite{bennett03} and references therein), and will continue to
furnish invaluable data in the years to come.  As the data volume and
precision demanded of these observations increases (for example, Bond
et al.~\cite{bond99}), the complexity of the analysis methods required
to deal with the data increases also.  Map-making -- the process of
turning time-ordered scan data into an image of the sky -- is an
example of a crucial step whose technical complexity has grown
significantly.  This is particularly so in the case of total power
measurements, where one removes signal drifts due to $1/f$-spectrum
noise in the map-making step.  If left unchecked, these drifts leave
stripes in the final map with amplitudes greater than the cosmic
signal, potentially compromising the scientific goals of a precision
instrument such as Planck.  For example, in the simulations presented
below, the magnitude of the striping signal (estimated from the rms
difference between a simple coadded map and the output map from one of
our codes) was 336~\muK, more than three times the output map's
residual rms error of $\sim$100~\muK\ due to white detector noise (see
Table~5).  (These numbers are for 1\farcm7 pixels and four polarized
detectors.  For 5\arcmin\ pixels, corresponding to the resolution of
the detectors, and for the full set of twelve detectors, the residual
error is $\sim$15~\muK\ for the temperature map.)  In other words,
map-making effectively removed striping with three times the target
sensitivity.  Proper map-making is thus crucial to mission objectives.

Planck\footnote{http://www.esa.int/science/planck/}, to be launched in
2008, will be the third-generation satellite dedicated to observations
of CMB anisotropies.  Its primary objective is to measure the
temperature anisotropies to the cosmic variance limit out to
multipoles $l>2000$; other scientific goals include detailed
measurements of the polarized power spectrum, the extraction of
catalogs of galaxy clusters and extragalactic sources, searches for
non-Gaussianity, and in-depth studies of the Galaxy.  To achieve these
goals, Planck will image the sky in nine frequency bands, with
resolution and sensitivity in the CMB-dominated bands of 5--15\arcmin\
and 5--10~\muK, respectively.  Crucial to the success of the mission
is the production of sky maps approaching the instrumental white-noise
limit; drifts and artifacts must be removed and the noise properties
well-understood. The formidable challenge of doing this for maps
containing millions of pixels lies at the heart of the effort of one
of the Planck working groups, the CTP Working Group.

It is important to develop and test well before launch efficient
algorithms for inclusion in the data analysis pipeline.  Besides
preparing the pipeline, this helps to identify potential sources of
systematic error and inform mission operations (e.g., scanning
strategy).  It also allows us to better quantify the mission's
expected scientific output.

In this paper, we evaluate a suite of map-making techniques using
simulations of several channels of the Planck High Frequency
Instrument (HFI) 217\,GHz time-ordered data (TOD).  The simulations
model non-white noise and primary CMB temperature and polarization
anisotropies.  The suite of methods includes both destriping and
optimal map-making algorithms.  We gauge the quality of our recovered
temperature and polarization maps by looking at rms pixel residuals,
and the residual power spectrum.  This gives us an evaluation through
second order statistical measures of noise and artifact residuals.
The complexity of the problem requires (at this stage) that we impose
a number of simplifying assumptions; these are clearly spelled out in
the text and are the focus of on-going work.  We emphasize that the
ability to produce the maps shown here is a notable achievement
requiring intensive computation.

\subsection{Planck Scanning Strategy}

Planck will make its observations from the 2nd Earth-Sun Lagrange
point, approximately $1.5 \times 10^6~\mathrm{km}$ from the Earth.
The satellite is spin-stabilized, and during science observations it
will rotate on its axis once per minute.  The telescope points at
angle of 85 degrees to the spin axis, so the detectors follow small
circles on the sky.  The satellite will perform a repointing manouvre
once per hour to keep the spin axis close to the anti-solar direction.
Thus during the one hour periods between manouvres the detectors will
make repeated observations of the same ``rings'' on the sky.  Some of
the algorithms presented in this paper can take advantage of these
repeated observations to reduce the computational burden.

Within the constraints imposed by Planck's design, there is freedom to
choose the precise pointings of the spin axis to optimize the
scientific returns of the mission.  The choice of spin axis pointings
-- the scanning strategy -- is one of the factors we examine in this
paper.

\subsection{Planck Science Goals for Sky Maps}

Planck is designed to image the sky at nine frequencies from 30 to
857\,GHz, with angular resolution from 33 to 5\,arcmin.  The raw
sensitivity is sufficient that inferences about the underlying
distribution of fluctuations on the sky should be limited not by
noise, but rather by cosmic variance.  To achieve this state,
systematic errors and processing artifacts must be controlled to
microkelvin levels.  It is the latter challenge that we are addressing
in this paper and its predecessor (Poutanen et al.~\cite{poutanen06}).

\section{Map-Making Algorithms}

Maps of the Cosmic Microwave Background (CMB) sky signal are derived
from long time series of data.  These data are generally collected by
telescopes fitted with detectors based on either HEMT amplifiers or
bolometers.  Both detection systems have noise characterized by a
power spectrum that rises at low temporal frequencies (often referred
to generically as ``$1/f$ noise'', even though this is only
approximate at best).  These systems are better-suited to differential
measurements, where a short period of time lapses between two
measurements being differenced, than to absolute ones.  To achieve
this, such CMB telescopes are scanned rapidly across the sky, with the
scanning pattern crossing over itself multiple times to enable
determination of the low frequency noise in the time stream.

\subsection{The Problem}

We assume that the data $\vd$ recorded from a detector can be written
as the sum of two contributions, one from the sky signal $\vs$ and the
other from the noise $\vn$ in the detection chain
\begin{equation}
  \vd = \vA \vs + \vn,
\end{equation}
where the pointing matrix $\vA$ describes the path of the detector
across the sky. If we have data from more than one detector, we can
treat all of the data together by concatenating their data vectors and
pointing matrices. The sky signal is represented by a vector with
discrete entries which are the pixels in a map. This implies that the
signal has a constant value across each pixel, so the pixels of the
map must be smaller than instrumental beam by a sufficient amount to
satisfy this assumption. If the detectors only measure the intensity
of the sky emission, then the signal vector contains one entry in each
pixel $p$, giving the value of the I Stokes parameter, $s_{pI}$. If,
however, the detectors also measure linear polarization, then the
signal vector must contain three entries for the I, Q, and U Stokes
parameters in pixel $p$, $s_{pI}$, $s_{pQ}$, and $s_{pU}$.

If at time-index $t$ an unpolarized detector is pointing at pixel $p$,
then the datum recorded from the detector is
\begin{equation}
  d_t = s_{pI} + n_t,
  \label{eqn:sample_unpol}
\end{equation}
where $n_t$ is the noise contribution to the datum. If the detector is
sensitive to polarization, the datum recorded is
\begin{equation}
  d_t = s_{pI} + s_{pQ} \cos(2\chi_t) + s_{pU} \sin(2\chi_t) + n_t,
  \label{eqn:sample_pol}
\end{equation}
where $\chi_t$ is the angle of the detector's polarization direction
with respect to the polarization basis in that pixel. It can be seen
from (\ref{eqn:sample_unpol}) and (\ref{eqn:sample_pol}) that the
pointing matrix $\vA$ is very sparse. Each row contains only one or
three non-zero entries, depending on whether the detector is
polarized.

The assumptions above imply that there is no attempt to deconvolve the
instrumental beam during the map-making process. If the beam is
symmetrical, then the resulting map will be smoothed with the same
beam. If instead the beam is asymmetrical, then the effective
smoothing of the map will vary with position. The smoothing at a
particular position will depend on the orientations in which the
detector has passed over that point in the sky. If the detector visits
all points on the sky in all orientations, then the effective
smoothing of the map will be given by a symmetrized version of the
beam~(Wu et al.~\cite{wu01}). This is not the case for the Planck
scanning strategy which is highly inhomogeneous.

It is assumed that the statistical properties of the noise $\vn$ are
known \textit{a priori}. It is possible to perform a joint estimation
of the signal and the noise properties, but this is not addressed in
this paper.  Here we assume that the noise is Gaussian with zero mean
and covariance $\vN$,
\begin{eqnarray}
  \left\langle \vn\right\rangle & = & 0,
  \label{eqn:npmean}\\
  \left\langle \vn\vn\tr\right\rangle & = & \vN,
  \label{eqn:npcov}
\end{eqnarray}
where $\left\langle\cdot\right\rangle$ denotes the average over noise
realisations. It is usually assumed that the noise properties are
stationary or piece-wise stationary in the time domain. Because of its
stationarity the noise can be described by a correlation function in
the time domain or equivalently by a power spectrum in the (temporal)
frequency domain. Thus each stationary block of the noise covariance
matrix is a symmetric Toeplitz matrix~(Golub \& van
Loan~\cite{golub96}).

\subsection{Optimal Solution}

The optimal solution for the map is given by the maximum-likelihood
estimate $\vshat$ which is obtained by solving the generalized
least-squares (GLS) equation
\begin{equation}
  \left(\vA\tr\vN^{-1}\vA\right) \vshat = \vA\tr\vN^{-1}\vd.
  \label{eqn:optmm}
\end{equation}
The term in brackets in the left-hand side is an $\Npix\times\Npix$
matrix, where $\Npix$ is the number of pixels in the map. It is the
inverse of the pixel-pixel covariance matrix of the resulting map,
\begin{equation}
  \vS = \left(\vA\tr\vN\inv\vA\right)\inv.
\end{equation}

For Planck-resolution maps, it is impossible in practice to calculate
$\vS\inv$, let alone invert it, so (\ref{eqn:optmm}) is solved using
iterative methods. The implementations of optimal map-making tested in
this paper all use the preconditioned conjugate gradient (PCG)
method~(Golub \& van Loan~\cite{golub96}) to find the solution. In each
iteration of the PCG method, it it necessary to apply $\vS\inv =
\vA\tr\vN\inv\vA$ to a vector. This can be achived by using its
factors in the following procedure:
\begin{enumerate}
\item Use the pointing matrix, $\vA$, to project the map-domain vector
into the time domain.
\item Apply the inverse noise covariance matrix $\vN\inv$ to the
time-domain vector.
\item Use the transpose of the pointing matrix $\vA\tr$ to map the
time-domain vector back into the map domain.
\end{enumerate}
Since the pointing matrix is sparse, steps (1) and (3) can be achieved
very efficiently in $O(N_{\mathrm{tod}})$ operations, where
$N_{\mathrm{tod}}$ is the number of TOD samples. Step (2) can be
applied by taking advantage of the piecewise-stationary properties of
the noise. The stationary blocks of the noise covariance matrix $\vN$
are symmetric Toeplitz matrices, and it is assumed that they are large
enough so that their inverses are well-approximated as Toeplitz
matrices too~(Stompor et al.~\cite{stompor02}). This being so, the
blocks can be applied using a convolution that is evaluted using a
fast Fourier transform (FFT). The computational complexity of this
step is $O(N_{\mathrm{tod}}\log N_{\mathrm{tod}})$. In practice, the
convolution kernel is truncated to a length $\Ntau$ where the
long-range correlations are dying away, so the computational
complexity is reduced to $O(N_{\mathrm{tod}}\log\Ntau)$.  The
convolution is applied using the overlap-add or overlap-save
methods~(Press et al.~\cite{press92}). Since the FFT is the most
time-consuming of these operations, all of the optimal map-making
codes in this paper use the highly-performant and portable FFTw
library~(Frigo \& Johnson~\cite{frigo98}).

The preconditioner used in the PCG solution is obtained by
approximating the noise covariance matrix $\vN$ as diagonal,
\begin{equation}
  \vN \sim \vNu = \mathrm{diag}(\ldots, \sigma^2_t, \ldots)
  \label{eqn:ndiag}
\end{equation}
where $\sigma_t$ is the noise standard deviation at time index
$t$. Thus the preconditioner matrix
\begin{equation}
  \vM = \vA\tr\vNu\inv\vA
  \label{eqn:precond}
\end{equation}
is diagonal in the unpolarized case and block diagonal in the
polarized case.

\subsection{Approximate Solution: The Destriping Approach}

In calculating the optimal solution to the map-making problem, many
computationally-expensive operations are required on data in the time
domain. If these could be reduced or eliminated by making
approximations of the noise, then the map-making process could be made
faster.

In the destriping approach (Burigana et al.~\cite{burigana97};
Delabrouille \cite{delabrouille98}; Maino et al.~\cite{maino99},
\cite{maino02}; Revenu et al.~\cite{revenu00}; Keih\"{a}nen et
al.~\cite{keihanen04}), the noise is divided into a low-frequency
component represented by a number of basis functions multiplied by
coefficients and a high-frequency part which is uncorrelated,
\begin{equation}
  \vn = \vB\vx + \vnu.
\label{eqn:noisetod}
\end{equation}
The coefficients $\vx$ are multiplied the matrix $\vB$ containing the
basis functions which are usually assumed to be a series of
offsets. The noise covariance matrix~(\ref{eqn:ndiag}) is diagonal,
though its diagonal elements need not be equal. The data are now given
by
\begin{equation}
  \vd = \vA\vs + \vB\vx + \vnu.
  \label{eqn:dmod}
\end{equation}
It is thus possible to write the map-making problem in a
maximum-likelihood form (Keih\"{a}nen et al.~\cite{keihanen04}),
where the parameters to be solved for are $\vs$ and $\vx$. The
maximum-likelihood coefficients $\vxhat$ can be found from the
TOD by solving
\begin{equation}
  \left(\vB\tr\vNu\inv\vec{Z}\vB\right)\vxhat
  = \vB\tr\vNu\inv\vec{Z}\vd,
  \label{eqn:offsoln}
\end{equation}
where
\begin{equation}
  \vec{Z} = \vec{I}-\vA\left(\vA\tr\vNu\inv\vA\right)\inv\vA\tr\vNu\inv.
  \label{eqn:zmat}
\end{equation}
The GLS equation (\ref{eqn:offsoln}) is solved using the conjugate
gradient method with or without a preconditioner. The offsets are then
subtracted from the TOD, which has the effect of ``whitening'' the
noise, so the map can be obtained by binning the resulting TOD
\begin{equation}
  \vshat = \left(\vA\tr\vNu\inv\vA\right)\inv
  \vA\tr\vNu\inv\left(\vd-\vB\vxhat\right).
  \label{eqn:dsmm}
\end{equation}

It is possible to incorporate prior information about the offsets in
the destriping process (Keih\"{a}nen et al.~\cite{keihanen05}). The
drifts in the TOD are caused by the low-frequency part of the noise
spectrum. The prior distribution of the noise is assumed to be
Gaussian with zero mean~(\ref{eqn:npmean}), so the prior on the
offsets must be Gaussian with a zero mean too. Thus the prior
distribution of the offsets can be summarized by a covariance matrix
$\vP$ which is related to the prior noise covariance matrix
(\ref{eqn:npcov}). Incorporating this into the solution for the
offsets, (\ref{eqn:offsoln}) becomes
\begin{equation}
  \left(\vB\tr\vNu\inv\vec{Z}\vB+\vP\inv\right)\vxhat
  = \vB\tr\vNu\inv\vec{Z}\vd.
  \label{eqn:offsoln_prior}
\end{equation}
The prior covariance of the offsets can be derived from the properties
of the noise spectrum or it can be estimated from the data.

\subsection{Degenerate Pixels}
\label{sec:deg_pixs}

If the map being made is polarized, then it is possible that some
pixels have not been observed by detectors in a sufficient number of
orientations to well constrain all of the I, Q, and U Stokes
parameters. If this is the case, then the sky signal estimate of all
three components may not be well-defined, and the corresponding
map-making equations are ill-conditioned. A procedure for
eliminating the badly-observed pixels is thus required in order to
solve the map-making equations in a general case.

Such a procedure can be based on the preconditioner matrix, $\vM$
(\ref{eqn:precond}).  The matrix plays a crucial role in both the
optimal and destriper approaches, and in the latter case, directly
determines the numerical stability of the sky map computations
(\ref{eqn:dsmm}).  In the optimal case, the (numerically stable)
invertibility of the matrix $\vM$ is a necessary (and for many
scanning strategies, nearly sufficient\footnote{The cases when this is
not a sufficient condition are, for example, the ones when all the
observations falling into a given pixel are nearly 100\% correlated.
Such cases can be readily avoided if each pixel on the sky is
revisited multiple times on different time scales. The additional
pixel excision criterion could then just read: $\Delta t >
1/f_{knee}$, with $\Delta t$ denoting the time interval between the
first and the last pixel observation.  Note that for the scanning
strategies and the characteristics of the instrument considered here,
such degenerate cases are unlikely to happen, as whenever the pixel is
observed from mulitple, sufficiently different directions, that
implies multiple visits of the pixel on a time scales comparable to or
longer than the noise correlations.}) condition for successful
calculation of the map.

Thus the preconditioner matrix can provide a method common to the
optimal and destriping approaches for eliminating badly-observed
pixels. The preconditioner matrix is block-diagonal with $3\times3$
blocks that correspond to the approximate covariance of the I, Q, and U
Stokes parameters of individual pixels.  Using (\ref{eqn:sample_pol})
and (\ref{eqn:ndiag}), it can be seen that the block of the matrix
corresponding to pixel $p$ is given by
\begin{equation}
  \vM_p = \sum_{d\in {\rm dets}}
  \sum_{t\in p} \frac{1}{\sigma_{d,\,t}^2}
  \left[
    \begin{array}{ccc}
      1 & \cos(2\chi_t) & \sin(2\chi_t) \\
      \cos(2\chi_t) & \cos^2(2\chi_t) & \cos(2\chi_t)\sin(2\chi_t) \\
      \sin(2\chi_t) & \cos(2\chi_t)\sin(2\chi_t) & \sin^2(2\chi_t)
    \end{array}
    \right],
\end{equation}
where the inner sum is over time-samples $t$ of the detector $d$
falling in the pixel $p$, and the outer one is over the detectors
included in the data set.  If the matrix is not invertible, then it is
impossible to constrain the Stokes parameters for this pixel.  We test
for the matrix invertibility using its condition number, $\kappa_p$,
(Golub \& van Loan~\cite{golub96}), defined as,
\begin{equation}
  \kappa_p \equiv \kappa\left(\vM_p\right) = {\displaystyle \max_{\rho_i
      \in \rho\left( \vM_p\right)} \left( \rho_i\right) \over \displaystyle
    \min_{\rho_i \in \rho\left( \vM_p\right)} \left( \rho_i\right)},
\end{equation}
where $\rho( \vM_p)$ denotes the eigenvalues $\{\rho_i\}$ of the
matrix $\vM_p$.  The condition number is thus calculated using an
eigenvalue decomposition routine (Golub \& van Loan~\cite{golub96};
Press et al.~\cite{press92}).  If it is larger than a chosen
threshold, then the Stokes parameters are considered to be
ill-constrained and the corresponding pixel is eliminated from the
map. (Note that the lower bound on the value of the condition number
is in our case equal to 2, corresponding to the perfectly isotropic
observations of a given pixel.)

Some of the codes presented in this paper implement a post-processing
phase in which further pixels are eliminated after the
map-making. This step uses either a lower value of the block condition
number threshold or an additional criterion using the absolute values
of the block eigenvalues or diagonal elements.  The pixels eliminated
in this way are not so poorly-conditioned as to make the solution
impossible, but they tend to produce large noise residuals and thus
are undesirable in the final map.

Note also that for the optimal approach removing pixels prior to
map-making needs to be done with care in order not to affect the
continuity of the time ordered data (Stompor et
al.~\cite{stompor02}). In the GLS runs described here, we excised a
pixel by setting the sky signal at the pixel to zero, and introducing
a fictitious extra pixel with zero sky signal observed instead. In
this way, we preserved the noise stationarity across the entire
segments of the data, without a need for involved time stream
processing.  None of these complications is relevant for the destriper
algorithm.  In the destriping runs presented here, samples
corresponding to the pixel to be removed are simply dropped from the
map-making.

\section{Map-Making Implementations}

Six implementations of the two basic algorithms described in Section~2
were tested on simulated Planck data. Three are based on optimal
map-making (MADmap, MapCUMBA, and ROMA), and three are based on
destriping (Springtide, Polar, and MADAM).  Common features of the
codes are described in Section~2.  In this section, we highlight
implementation details and differences.

\subsection{ROMA}

ROMA (ROMA Optimal Map-Making) is a Fortran 95 parallel (MPI)
implementation of an iterative optimal GLS map-making algorithm.  The
first implementation of ROMA (Natoli et al.~\cite{natoli01}) showed
that it is feasible to solve the optimal map-making equations
(\ref{eqn:optmm}) for a Planck-sized dataset using an FFT-based, PCG
iterative solver.  Over time, ROMA has evolved into a multichannel,
polarization-capable code, which includes an iterative noise
estimation stage.  ROMA has been integrated at the Planck LFI Data
Processing Center; it has also been used to analyze data from the last
(2003) Antarctic flight of BOOMERanG.  A detailed discussion can be
found in (de~Gasperis et al.~\cite{degasperis05}).

The noise was considered stationary over the whole timeline, and its
power spectral density was assumed known \textit{a priori}. The
instrumental beam is assumed to be axisymmetric and common to $I$,
$Q$, and $U$.  This means that the code solves for a beam-smeared $(I,
Q, U)$ set of maps.  The condition number of each block of the
preconditioner matrix (\ref{eqn:precond}) is checked and, if deemed
satisfactory, the block is inverted and the result used for
preconditioning.  The convolution is performed using the overlap save
method (Press et al.~\cite{press92}), exploiting the fact that the the
noise correlation function is truncated after a lag (equal to 16385
samples.  Double precision is used throughout.

\subsection{MADmap}

MADmap, part of the MADCAP3 package, is a massively parallel
implementation of the map-making equations (\ref{eqn:optmm}); see
Borrill et al.~(\cite{borrill06}) for a detailed description.
Communications between processors are handled by MPI, and data are
distributed over the memory associated with each processor.

The major data objects involved in the calculation are distributed
over the memory associated with each processor in a balanced way.  All
time domain data, including the time stream vector, the pointing
matrix, and the correlation functions that define the inverse
time-time noise matrix, are distributed with an even partition of the
data over the processors so that every processor stores the data
associated with an equal number of contiguous time samples.  A single
processor may store time stream data that belong to more than one data
set (a data set is the data recorded by a single detector in the
simulations described in this paper) if the processor stores the last
samples from one data set and the first samples from the next.

In calculating the pixel-pixel noise correlation matrix with the PCG
algorithm, five pixel-domain vectors must be stored.  These vectors
are distributed over the processors so that if a pixel is observed
during the time interval associated with a given processor, then the
vector elements associated with that pixel are stored in the
processor's memory.  This does not represent an even partition of the
pixel domain, but it does limit the size of the pixel vectors stored
on a given processor to be less than the size of the pointing matrix
stored on the processor (with most experimental scanning strategies it
will be significantly smaller).  For the Planck scanning strategy in
particular, the memory footprint required for the pixel vectors will
be approximately 60 times smaller than the pointing matrix memory
footprint.

The pointing matrix is stored as a sparse matrix in zero-compressed
row major order; it can take on a variety of forms to reflect
different models of the observation.  If there are $N_z$ non-zeros per
row, then $\vA$ takes $O(N_z N_t)$ storage space and $O(N_z N_t)$
calculations to operate with, where $N_t$ is the number of time
samples.  MADmap allows for an arbitrary pointing matrix, and this
flexibility allows for very complicated time stream models.  By using
an arbitrary pointing matrix we can remove parasitic sky-asynchronous
signals, calibrate from a known signal, pixelate a systematic effect
and remove it, and so forth.

The inverse time-time noise correlation matrix operator has a complex
implementation (Borrill et al.~\cite{borrill06}). The computational cost
of the $\vN\inv$ operator is $O(N_t \ln( N_\tau ) )$ and the storage
space required is $O(N_s N_\tau)$, where $N_t$ is the number of time
samples, $N_\tau$ is the length of the correlation function, and $N_s$
is the number of stationary intervals.  It is important to note that
$\vN\inv$ can be arbitrary within the structural definition outlined
above.  This includes the possibility for each stationary interval to
have a different noise correlation function associated with it, and
these noise correlation functions can take any form.

MADmap has the functionality to read an arbitrary sparse
preconditioner matrix and use this precomputed data product as the
preconditioner.  If no preconditioner is provided, then MADmap uses a
diagonal preconditioner composed of the inverse of the diagonal
elements of $\vA\tr\vA$.

MADmap uses the M3 library to access CMB data.  Applications that use
the M3 library for reading data can be daisy chained, so that the
output of one application can be used as input for a subsequent
application.  (A suite of massively parallel applications using the M3
library is detailed in Borrill et al.~(\cite{borrill06})).  This
functionality is afforded by the combination of a versatile XML-based
data model description and the ability of the M3 library to read data
stored a variety of file formats while translating the data to conform
to a single application programmer interface.

Use of the M3 library also gives MADmap a comprehensive and versatile
user interface designed specifically to enable the analysis of
simulated data.  For each detector and for each six day period there
are three files, containing the pointing information, the CMB signal
time stream, and the detector noise time stream.  These 732 files were
organized in the M3 XML document describing the simulation, and the M3
library is able to add the signal and the noise time streams in a
weighted fashion as specified by the user.

\subsection{MapCUMBA}

MapCUMBA is a set of codes dealing with the analysis and map-making of
polarized CMB data. The current version (2.1) has been modified from
previous versions (Dor\'{e} et al.~\cite{dore01}) to solve the GLS
map-making equation (\ref{eqn:optmm}) by the preconditioned conjugate
gradient (PCG) method. It can now process polarized data from multiple
detectors, filling the gaps in the data stream due to cosmic ray hits
or data acquisition drop-out with a constrained noise realization. The
noise properties, assumed piece-wise stationary, can be estimated from
the TOD through an iterative process of map-making and noise
extraction. The codes are written in Fortran 90 and are parallelized
for distributed memory architecture using the message passing
interface (MPI).  They have been applied to the analysis of the 2003
flight of Boomerang.

In the current analysis, the instrumental noise is assumed to have a
known power spectrum and to be statistically stationary over the whole
flight.  We assume its time-time correlations to be negligible beyond
5.4~min, which corresponds to $\Ntau = 65536$ samples.  This latter
number has been chosen to be a power of two to speed up the FFT
operations described below, but could be set to any other value larger
than the noise correlation length.

The convolution is applied using an overlap-add method (Press et
al.~\cite{press92}).  This approach allows for a simple
parallelization of the code, reducing the computation and memory load
of each CPU.  Since the number of pieces created is generally larger
than the number of CPUs, and they have a predictable length (they are
all equal here), load balancing between CPUs is fairly simple.

In order to apply the PCG techniques, the matrix to be inverted
$\vA\tr \vN\inv \vA$ is approximated by $\vM = \sum_c \vA\tr_c
\sigma^{-2}_c \vA_c$, where $c$ is a data stream of a given detector
for which the intrumental noise is assumed stationary with
rms~$\sigma_c$. $\vA_c$ is the pointing matrix of the detector during
that stretch of time, and the sum is done over all such pieces for all
detectors.  For polarization, if the condition number (ratio of the
largest to the smallest eigenvalues) of a given block of $\vM$ is
small enough, the block is inverted to provide a local preconditioner.
Otherwise the pixel is degenerate and flagged as unobserved.

Making maps of the Planck HFI data with a optimal method and a PCG
algorithm poses a double challenge in terms of data volume.  The full
data stream must be kept in memory or re-read regularly from disk to
allow for the filtering operations described earlier, and the PCG
algorithm requires five intermediate maps and the preconditioner
$\vM\inv$, each covering the same sky area as the final map with the
same resolution and number of fields, to be kept in memory at the same
time.  For the full-sky HFI applications considered here, each
unpolarized map with $0\farcm8$ pixels ($\Nside = 4096$) takes
$1.6~\mathrm{GB}$ in double precision, while a polarized map with
$1\farcm7$ pixels ($\Nside = 2048$) takes $1.2~\mathrm{GB}$.  The maps
used in the PCG solver are therefore broken into $N_\mathrm{cpu}$
non-overlapping pieces of almost equal number of pixels, and each
piece is assigned to one CPU.  This scheme simplifies the operations
such as scalar products of maps, and application of the preconditioner
performed during the PCG iterations, and ensures a perfect load
balancing independently of the scanning strategy.

\subsection{Polar}

Polar (Keih\"{a}nen et al.~\cite{keihanen06}) is based on the
destriping approach.  We assume no a priori knowledge of the baseline
amplitudes, assigning uniform prior probability to $\vx$. The
baseline amplitude vector $\vx$ is solved from the TOD using
(\ref{eqn:offsoln}), without preconditioning.

In the present study we use a baseline length of 1\,min, the nominal
spin period of the Planck spacecraft. The number of baselines is then
$N_b = N_\mathrm{det}\times366\times24\times60$, where
$N_\mathrm{det}$ is the number of detectors used.  We have found that
1-min baselines produce lower noise in the maps than longer baselines
(Keih\"{a}nen et al.~\cite{keihanen06}).  Note that no coaddition of
the data was performed, unlike in Keih\"{a}nen et
al.~(\cite{keihanen04}) and Poutanen et al.~(\cite{poutanen06}),
where the use of idealized pointing combined with the Planck scanning
strategy made it easy to compress the TOD into smaller size by
coadding repeated scans over the same sky circles.

Solving for the baselines and the final map-making step involve
multiplication by the inverse of the matrix $\vA\tr\,\vN_u\,\vA$
(\ref{eqn:precond}). This matrix has been discussed in more detail in
Section~\ref{sec:deg_pixs}. As noted there, that in order to ensure
its invertibility, we need to remove the degenerate pixels.  The Polar
code has two options for the criteria to determine those pixels. One
is a determinant criterion, defined as $\det \equiv \sigma^6\det
\vM_p$, where the same white noise level $\sigma_i = \sigma$ is
assumed for each detector and sample and $\vM_p$ denotes the p-th
diagonal block of the matrix $\vA\tr\,\vN_u\,\vA$. The second
criterion is the condition number requirement introduced in
Section~\ref{sec:deg_pixs}. The pixels, whose determinant or
reciprocal condition number is less than some predefined value, will
be rejected.

The code is written in Fortran 90, and parallelized using MPI.  The
full data stream is kept in memory, leading to a large memory
requirement.  Each processor handles a section of the TOD.  In this
study the sections were six days long.  The total number of sections
and processors required, therefore, was $N_\mathrm{det}\times61$.
Because of the memory requirement, however, 512 (1536) processors were
reserved for $N_\mathrm{det} = 4$ (12).  The map is divided into
submaps (in this case 1024 pixels each).  Each processor handles a
number of submaps.  Processors handling TOD and processors handling
maps (the same processors may do both tasks) communicate by sending
these submaps to each other.

\subsection{MADAM}

The MADAM map-making method is described in Keih\"{a}nen et
al.~(\cite{keihanen05}).  The method is based on the destriping
technique, but also uses information on the noise spectrum.

The first implementation of MADAM was serial, and included no
polarization.  It allowed fitting several basis functions, such as
Fourier components, and also allowed for coadding of the TOD.  For the
purpose of this study we wrote a new parallel implementation of the
method, capable of handling larger data sets, and specially designed
for uncoadded data.  It allows fitting of uniform baselines only,
which simplifies the code somewhat, but is able to handle polarization
data.

As in Polar, the low-frequency part of the noise is represented by a
sequence of constant baselines, and the remaining part of the noise is
treated as white noise. The baseline amplitude vector $\vx$ is solved
from equation (\ref{eqn:offsoln_prior}). The covariance matrix $\vP$
of the baseline amplitudes (\ref{eqn:offsoln_prior}) is computed from
the noise spectrum, assumed to be known. The final output map is
constructed from the cleaned TOD as in equation (\ref{eqn:dsmm}).

The code has some parts in common with Polar.  In particular, the
binning of TOD into the map and handling of degenerate pixels are
done in the same way. Also, the parallelization scheme and
distribution of data among processors are similar.  The main
difference between the codes is the term $\vP\inv\vx$ in
(\ref{eqn:offsoln_prior}), which allows MADAM to fit accurately much
shorter baselines than Polar and thus model the correlated noise
better. In the present study we used a baseline length of 4~s.  The
chosen baseline length represents a trade-off between accuracy and
computation time.

MADAM solves the linear system (\ref{eqn:offsoln_prior}) by conjugate
gradient iteration, with a preconditioner matrix
\begin{equation}
  \vM=(\vB\tr\vNu\inv\vB+\vP\inv).
  \label{precond_madam}
\end{equation}
The preconditioner is circulant and its multiplication with a
baseline vector can be handled by a Fourier technique, like the
term $\vP\inv\vx$.

The term $\vP\inv\vx$ is computed using the overlap-save method (Press
et al.~\cite{press92}).  It is calculated independently for each
detector.  The sequence of baselines is split into sections, whose
length in the present study was 6 or 3 days. Each section is padded at
both ends with baselines from the two neighbouring sections, wrapping
around at the ends.  Each section is then convolved with $\vP\inv$
using a FFT technique and cut back to the original length. The
convolution takes into account noise correlations over the length of
one data section.

Formulae for computing the prior offset covariance $\vP$ were given
in Keih\"{a}nen et al.~(\cite{keihanen05}).  The code assumes
stationary noise over the whole TOD sequence.

\subsection{Springtide}

Springtide (Ashdown~\cite{ashdown06}) is an implementation of
destriping that is optimized to enable medium-sized supercomputers or
clusters to analyse Planck-sized data sets.  The map-making
inplementations described above require all of the data and pointing
to be stored in memory at once, at large cost in memory.  For example,
to process the data from all 12 detectors of the HFI
$217~\mathrm{GHz}$ channel requires several terabytes of RAM.
Springtide reduces the memory requirements by dividing the map-making
into two stages.

The first stage takes advantage of the repeated observations of the
same ring during one pointing period to compress the data.  The
baselines that will be subtracted from the time-ordered data are taken
to be equal to the pointing period, during which the noise is assumed
to be white.  The data can therefore be compressed by making a map for
each pointing period for each detector -- a ``ring-map'' -- by simple
binning.  The low-frequency part of the noise, modelled by an offset
for each pointing period, is unaffected by the binning operation.  The
high-frequency noise in the resulting ring-maps is uncorrelated, with
an amplitude modulated by the hit-count in each of the ring-pixels.
Making ring-maps in this way compresses the TOD by a factor of 20--30,
depending on the amplitude of the nutation of the satellite.  The
ring-maps are made using the same pixelation that will be used to make
the final map to avoid any re-pixelation errors.

The second stage of the map-making is to combine the ring-maps into a
global map by destriping.  In each ring-map, the low-frequency noise
is modelled as an offset.  The maximum-likelihood solution for the
offsets is found using the conjugate gradient algorithm.  The
maximum-likelihood offsets are subtracted from the ring-maps, which
are then binned to make the global map.  An iterative procedure can be
used to estimate the low-frequency properties of the noise.

Springtide is written in Fortran 95 and is parallelized using MPI.

\section{Simulations}

\subsection{Assumptions}
\label{subsec:assumptions}

We assumed the following:

\begin{itemize}

\item Noise is known a priori.  (In an experiment it will have to be
determined from the data themselves.  Here we unrealistically avoid
the effects of both the error and the statistical uncertainty that
would exist in dealing with real data.)

\item Noise consists only of white noise and a low frequency
component, with no peaks or other features at high frequencies, and no
gaps in the noise spectrum (as would result from, e.g., microphonics
of slightly varying frequency).

\item Noise is pseudo-stationary and Gaussian. The data are divided
into 6-day pieces; noise is uncorrelated between pieces, but noise
properties are identical in all pieces.

\item The white noise level is $\sigma=1.280$\,mK for unpolarized and
$\sigma=1.811$\,mK for polarized timelines.

\item The knee frequency $f_{\mathrm{knee}} = 0.03~\mathrm{Hz}$,
minimum frequency $f_{\mathrm{min}} = 1.15\times10^{-5}~\mathrm{Hz}$,
and spectral slope $\alpha=-2$.

\item Noise is uncorrelated between detectors.

\item Beams are symmetric and Gaussian.

\item FWHM = 4\farcm7269, 4\farcm7470, 4\farcm7123, and 4\farcm7116
for 217-5a, 217-5b, 217-7a, and 217-7b, respectively.

\item There are no missing data in the time-stream

\item There are no foreground signals.

\end{itemize}

For each detector, measurements are taken at a rate of 200\,samples/s.
The detectors scan the sky along $\sim 85^\circ$-opening-angle
circles, at 1~rpm. The spin axis is changed once per hour by 2\farcm5.
Two scanning strategies were simulated:

\begin{itemize}

\item {\bf nominal scanning}, in which the spin axis follows the
ecliptic plane in the anti-Sun direction during the whole flight;

\item {\bf cycloidal scanning}, in which the spin axis follows a
cycloidal path 7\degr\ from the anti-Sun direction with 6-month
period.

\end{itemize}

For both scanning strategies, a small nutation of the spin axis was
included, along with a small change in the spin rate.  The nutation
amplitude and spin rate were changed every hour when the spin axis was
repointed, and then kept constant for an hour.  Their values were
selected randomly from truncated Gaussian distributions with 0\farcm5
rms and 2\arcmin\ maximum for nutation and 0\fdg12\,s\mo\ rms and
0\fdg3\,s\mo\ max for spin rate deviation from the nominal 1\,rpm.

\subsection{The CMB template}

The CMB reference sky was constructed to reproduce the large scale
anisotropy pattern observed by WMAP.  Specifically, we chose a
threshold in the angular domain, $l_{\mathrm{WMAP}}$, and used the
observed harmonic coefficients for $l\le l_{\mathrm{WMAP}}$ from the
internal linear combination (ILC) CMB
map\footnote{http://lambda.gsfc.nasa.gov/product/map/m$_{-}$products.cfm}
(Bennett et al. 2003) as described below.  To reduce the residual
effect of foregrounds and WMAP instrumental noise and systematics, we
chose the threshold to be much larger than the actual WMAP angular
resolution, i.e., $l_{\mathrm{WMAP}}=70$.  The CMB reference consists
of total intensity and $E$ mode polarization, but zero power on $B$,
and has been obtained as described below.  Extensive use was made of
HEALPix\footnote{http://healpix.jpl.nasa.gov/} routines.  The template
and the codes for producing it are publicly available as part of the
Planck Component Separation Working Group effort for building a
background and foreground reference sky\footnote{see
http://www.planck.fr/heading79.html for details}.

\subsubsection{$l\le l_{\mathrm{WMAP}}$}

For total intensity, we take directly the $a^{T,\mathrm{WMAP}}_{lm}$
from the outputs of the HEALPix anafast routine running on the WMAP
ILC map.  The coefficients for the $E$ polarization mode,
$a^{E}_{lm}$, were obtained as follows.  For all $m$, the average
amplitude of the component of $a^{E}_{lm}$ which is correlated with
$T$ is given by $C^{TE}_{l}/C^{T}_{l}$; the remaining component has
average amplitude $\sqrt{C^{E}_{l}-(C^{TE}_{l}/C^{T}_{l})C^{TE}_{l}}$.
Here by $C_{l}$s we mean those taken from the best fit theoretical
power spectrum to the WMAP, ACBAR, and CBI data, described below.  The
relative weights of the real and imaginary part of the $a^{E}_{lm}$
component which is correlated with $TE$ are set by
$a^{T,\mathrm{WMAP}}_{lm}$.  Those of the remaining part are chosen
randomly. The result is
\begin{equation}
\label{formula}
a^{E}_{lm}=
a^{T,\mathrm{WMAP}}_{lm}\cdot\frac{C^{TE}_{l}}{C^{T}_{l}}+
\left(\frac{x^{E}_{lm}+iy^{E}_{lm}}{\sqrt{2}}\right)
\sqrt{C^{E}_{l}-\frac{C^{TE}_{l}}{C^{T}_{l}}C^{TE}_{l}}\ ,
\end{equation}
where $x^{E}_{lm}$ and $y^{E}_{lm}$ are Gaussian distributed numbers
with zero mean and unit variance; the $\sqrt{2}$ as well as the
imaginary part are absent for $m=0$.

\subsubsection{$l> l_{\mathrm{WMAP}}$}

The harmonic coefficients on these scales were obtained by running the
synfast HEALPix routine to generate a template for $T$, $Q$, and $U$
out of the WMAP, ACBAR, and CBI best-fit theoretical CMB power
spectrum with no running spectral
index\footnote{http://lambda.gsfc.nasa.gov/product/map/lcdm.cfm},
specified by $h=0.71992$, $\omega_{b}=0.02238$,
$\omega_{CDM}=0.11061$, $\tau=0.11027$, $n_{s}=0.95820$, and $r=0$,
where $h$ is the Hubble constant in units of 100\,km\,s\mo\,Mpc\mo,
$\omega_{x}/h^{2}$ is the ratio between the density of the species $x$
and the critical one, $\tau$ is the optical depth at reionization,
$n_{s}$ is the scalar perturbation spectral index, and $r$ is the
ratio between the primordial tensor and scalar perturbation amplitude.

\subsection{Temperature}

An initial set of simulations for temperature only, summarized in
Table~\ref{tab:temp_cases}, was run with no noise, in order to test
our ability to reproduce exactly the input map.  These simulations
comprised one year of CMB temperature data from HFI~217-4, using the
cycloidal scanning strategy over an input temperature map generated at
HEALPix \Nside = 4096 resolution.  The pointing data were generated in
double precision and stored in either double or single precision (see
Table~\ref{tab:temp_cases}); other data were stored in single
precision.

\begin{table}[h]
\begin{center}
\begin{tabular}{c c c c}
\hline\hline
Case & Bolometer time constant & Pointing precision &
Output resolution / \Nside \\
\hline
1 & No  & Double & 4096 \\
2 & No  & Single & 4096 \\
3 & Yes & Double & 4096 \\
4 & No  & Double & 2048 \\
\hline
\end{tabular}
\caption{Inputs and outputs of the four temperature simulations
without noise, designed to test our ability to reproduce the input map
exactly.}
\label{tab:temp_cases}
\end{center}
\end{table}

Three main issues were identified which impacted the ability of the
map-making codes to reproduce the noiseless signal in the input maps:

\begin{itemize}

\item{Pointing precision -- Information about the satellite pointing
is specified in terms of $\theta$ and $\phi$, the co-polar and
azimuthal angles in ecliptic coordinates.  Initially, the Level-S
pipeline calculated the pointing angles in double precision but stored
them in single precision.  This truncation led a small fraction of the
pointings being assigned to the wrong pixel.  This mis-assignment
could only affect pointings that lay within the truncation error limit
of a pixel boundary, but the number of such pointings increases as the
pixel size is reduced.}

\item{Bolometric time constants -- Each time sample corresponds to the
integration of the signal along the scan for the duration of the
sampling period.  The Level-S pipeline simulates this by performing a
convolution of the input signal along the scan.  The time stamp
associated with each sample is the time when the sample is read by the
onboard electronics, at the end of the integration period.  In
addition to this integration, bolometers have a finite time impulse
response.  This leads to an additional smearing which is modeled in
the simulations by a convolution with a causal filter (implemented
through averaging over `fast samples'). The smearing of the signal due
to both of these effects will have to be taken into account in the
analysis. Ideally the filter response will be inferred from the data
themselves.  We conjecture that the ability to do so will depend on
the availability of bright point sources and the ratio of the
bolometric time constant to the sampling period.  Except for the
noiseless case 3, where this feature was included in the simulation,
to see its effect when not corrected for, we decided to switch it off
from the simulation pipeline for several reasons:

\begin{itemize}

\item{In practice, these effects will have to be included in the
time-stream pre-processing and noise estimation, not the map-making,
and this stage was not being considered here.}

\item{The effect of the bolometric time constants is coupled to the
effects of asymmetric beams and correlated noise.  Our current
exercise does not consider these effects, and it does not make sense
to consider the effect of bolometric time constants in isolation.}

\item{The bolometric time constants used in Level-S were not based on
actual hardware performance, which was not yet available, and
therefore were not likely to be representative of the actual Planck
instrument.}

\end{itemize}

}

\item{Output map resolution -- If the output map has
lower resolution than the input map there is a residual due to the
additional pixel smoothing.}

\end{itemize}

Once these issues were sorted out, a fifth temperature-only simulation
was run, the same as Case~1 but with noise added.

\subsection{Polarization}

We then ran simulations with polarization, Based on the lessons
learned in the temperature cases, all used double precision pointing,
no bolometer time constant, and output maps at the same resolution as
the input data.  Data were generated for four polarization-sensitive
bolometers, 217-5a, 217-5b, 217-7a \& 217-7b.  Input and output $I$,
$Q$, and $U$ maps were made at $\Nside = 2048$ to reduce memory
requirements during the simulation process.  Four cases were run: with
and without noise for both nominal and cycloidal scanning strategies.

\section{Results}
\label{sec:results}

\begin{figure*}
  \centering
  \includegraphics[height=16cm,angle=-90]{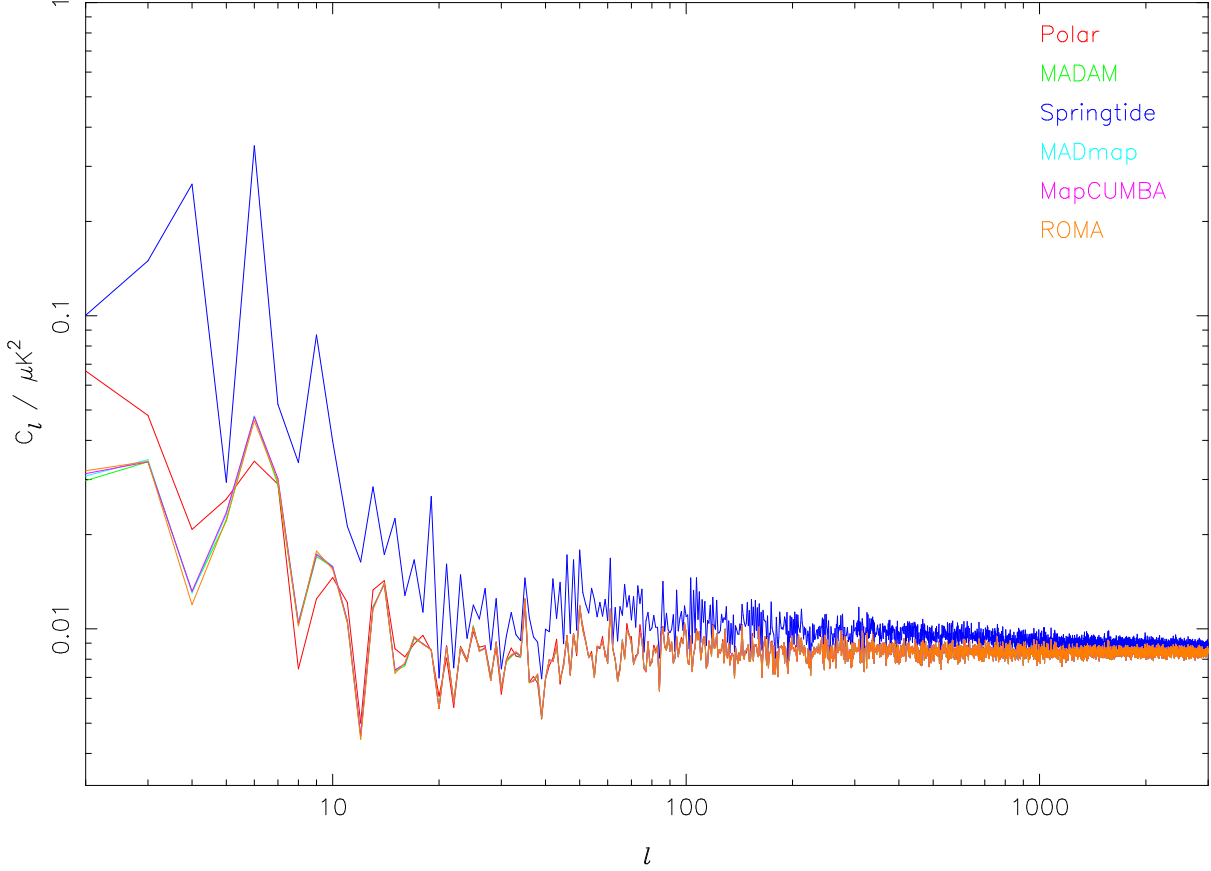}
  \caption{Spectrum of residuals from the temperature simulation with
    noise.}
\label{fig:ClT}
\end{figure*}

The output map was subtracted from a reference map at the same
resolution, and the mean, minimum, maximum, and root mean square
residual pixel values were calculated, together with the angular power
spectrum of the residual map.  The full results are available on the
Helsinki results
website\footnote{http://www.mrao.cam.ac.uk/$\sim$maja1/ctp/helsinki/}.
Tables~\ref{tab:Tno_noise} (no noise), \ref{tab:Tnoise} (with noise),
and Figure~\ref{fig:ClT} (with noise) summarize the results for the
single detector temperature-only runs for the four cases listed in
Table~\ref{tab:temp_cases}.

\begin{table}[h]
\begin{center}
\begin{tabular}{l c c c c}
\hline\hline
     & \multicolumn{4}{c}{RMS residual / \muK} \\ \cline{2-5}
Code              & Case 1 & Case 2 & Case 3 & Case 4 \\
\hline
Polar             & $0.00$                & $10.9$ & $0.0753$ & $1.85$ \\
MADAM             & $0.00$                & --     & --       & $1.85$ \\
Springtide        & $0.00$                & --     & --       & $1.85$ \\
MADmap$^{\ast}$   & $2.08\times 10^{-18}$ & $10.9$ & $0.0682$ & $1.85$ \\
MADmap$^{\dag}$   & $4.50\times 10^{-4}$  & --     & --       & --     \\
MapCUMBA$^{\ast}$ & $2.06\times 10^{-5}$  & $10.9$ & $0.0682$ & $1.85$ \\
ROMA$^{\ast}$     & $0.00$                & --     & --       & $1.85$ \\
ROMA$^{\dag}$     & $0.00$                & --     & --       & --     \\
\hline
\end{tabular}
\caption{RMS pixel residual for the four temperature simulations
without noise. For the optimal map-making codes ${}^{\ast}$ indicates
that noise correlations were ignored and ${}^{\dag}$ that they were
taken into account; see the text for further details.}
\label{tab:Tno_noise}
\end{center}
\end{table}

\begin{table}[h]
\begin{center}
\begin{tabular}{l c c c c}
\hline\hline
Code & RMS residual / \muK \\
\hline
Polar           & $379.12$ \\
MADAM           & $378.96$ \\
Springtide      & $401.19$ \\
MADmap          & $378.96$ \\
MapCUMBA        & $378.96$ \\
ROMA            & $378.96$ \\
\hline
\end{tabular}
\caption{RMS pixel residual for the temperature simulation with
noise.}
\label{tab:Tnoise}
\end{center}
\end{table}

In the noiseless cases, the optimal map-making implementations can be
run with two different assumptions about the noise covariance
matrix. The first is to assume that the prior information about the
noise is correct, even though no noise is included in the simulations.
The second is to ignore the prior information about the noise and to
assume it is white.  For the noiseless case the latter is closer to
the truth, but ignores the correlations that will be introduced when
noise is added to the simulations.  Tables~\ref{tab:Tno_noise} and
\ref{tab:Pno_noise} indicate which assumption was made.

From the no-noise results we see that simulation errors exceed the
numerical precision of the map-making codes and are recovered by all
six methods.  Residuals arising from the map-making itself are
extremely small, as can be seen from the Case~1 results in
Table~\ref{tab:Tno_noise}.  The much higher RMS of the codes in Case~2
illustrates the point made in Section~4.2 about the impact of pointing
precision on high resolution maps.  Case~3 shows that smoothing caused
by the bolometer time constant adds appreciable residuals.  (The
marginally higher rms for Polar in Case~3 is due to using a lower
pixel resolution ($N_{\mathrm{side}} = 2048$) in the ring offset
fit. This was a temporary fix for a problem an early version of the
code had with noiseless data. The TOD for Cases~2 and~3 were later
deleted to save disk space, so that they could not be redone with a
consistent use of the higher resolution, after the problem was
corrected.) Case~4 shows that the use of different resolutions for
input and output maps contributes much larger residuals than the
map-making itself. The most significant of these three sources or
error is the pointing precision, followed by the output map
resolution, and then the bolometer time constant.  It is worth noting
that the pointing precision error generates residuals with not only
the largest amplitude, but also the most structured angular power
spectrum.

In all cases the (pure) map-making residuals are much smaller than the
residual noise level, as shown by the results given in
Table~\ref{tab:Tnoise} for the simulations including instrumental
noise.  As discussed in Section~5.1 below, the residual RMS values in
Table~\ref{tab:Tnoise} are consistent with the expected white noise
level.  Figure~\ref{fig:ClT} displays the power spectrum of the noise
residuals.  All the methods tend towards the white noise plateau at
high $l$, and contain significant structure at low $l$.  Residual
noise from Springtide remains noticably higher than in the other
codes, because it works with ring baselines rather than the shorter
time-stream baselines used by the other destripers.  This greatly
reduces the resources required for the code (see below), but it
permits a larger amount of low-frequency noise to pass into the
reconstructed map.

\begin{figure*}
  \centering
  \includegraphics[width=16cm]{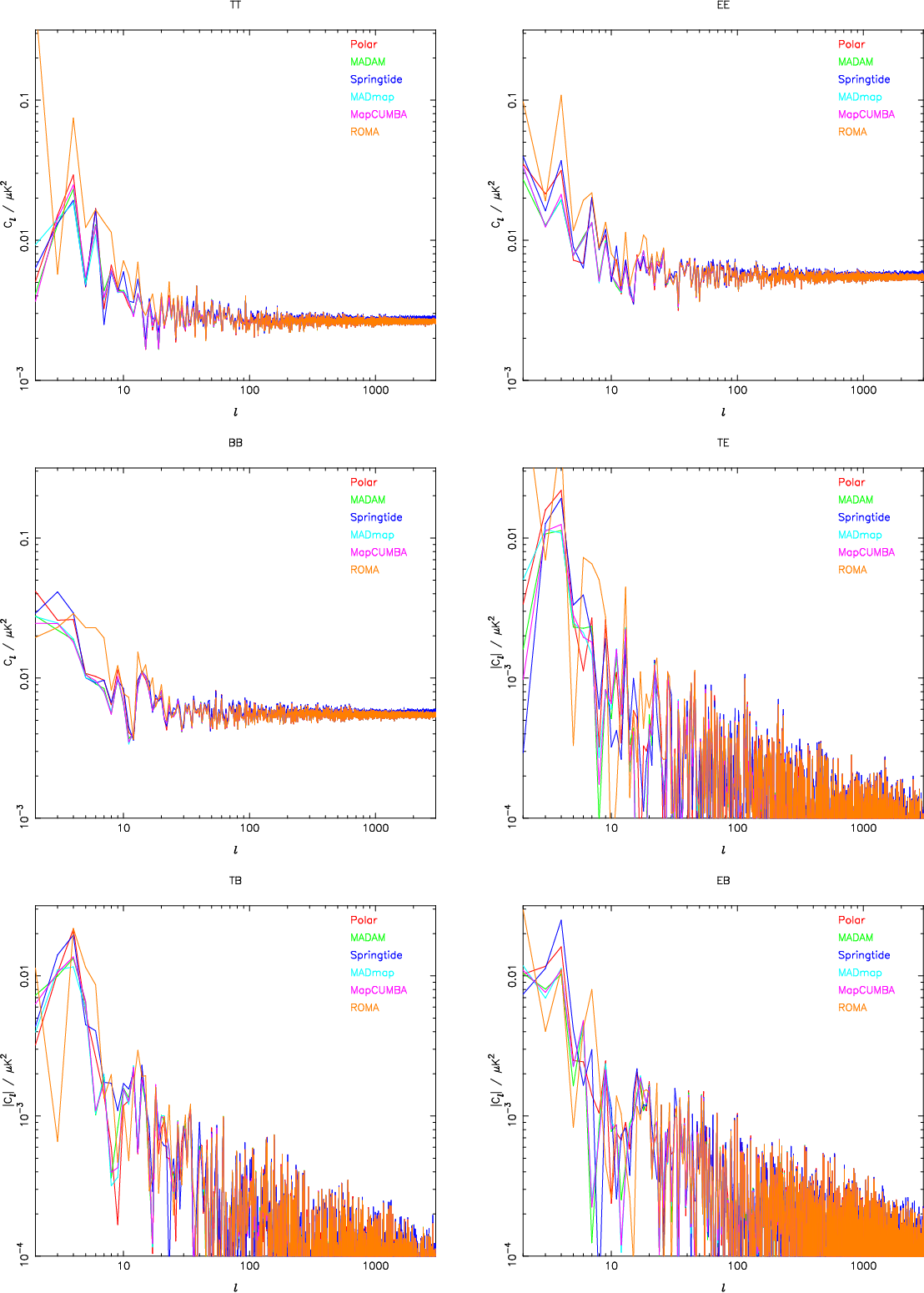}
  \caption{Spectra of residuals for the polarization simulations with
    noise in the cycloidal scanning strategy.}
  \label{fig:ClP_cycloidal}
\end{figure*}

\begin{figure*}
  \centering
  \includegraphics[width=16cm]{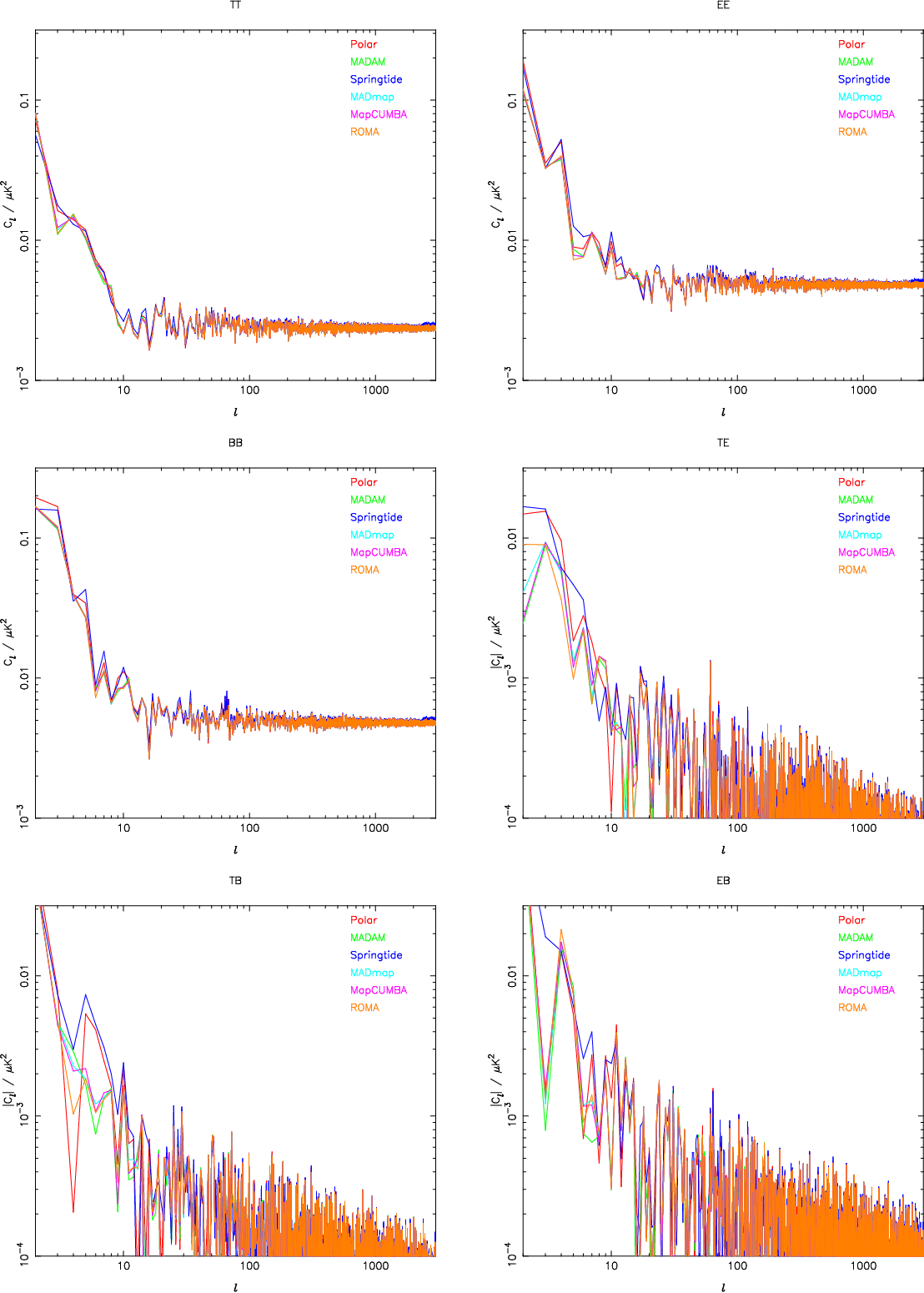}
  \caption{Spectra of residuals for the polarization simulations with
    noise in the nominal scanning strategy.}
  \label{fig:ClP_nominal}
\end{figure*}

Tables~\ref{tab:Pno_noise} (no noise) and \ref{tab:Pnoise} (with
noise) and Figures~\ref{fig:ClP_cycloidal} (cycloidal scanning) and
\ref{fig:ClP_nominal} (nominal scanning) show results for the
polarization simulations. The beam sizes of the four detectors differ
slightly; the reference map used to calculate the residuals was
constructed by averaging the four input maps.  Comparison of the two
tables demonstrates that once again residuals caused by the map-making
itself are negligible relative to the noise residuals.  We also note
that the U residuals in Table 4 are significantly higher than those on
Q, an effect that we have traced to differences in the FWHM of the
beams (uncorrected in our map-making). In fact, the difference in beam
FWHM for the 217-5ab pair, aligned at 45 degrees to the scan
direction, is larger than the difference for the 217-7ab pair, aligned
parallel/perpendicular to the scan direction.  Given these
orientations and the scanning pattern on the sky, the 217-5ab pair
primarily measures the U field (especially at the Ecliptic equator).
The larger beam difference for this pair thus causes a larger residual
in the U measurement.  The effect is, however, much smaller than the
noise residuals (Table 5). We discuss the import of these results on
scanning strategy in the Conclusion section.

\begin{table}[h]
\begin{center}
\begin{tabular}{l c c c c c c}
\hline\hline
     & \multicolumn{3}{c}{Cycloidal RMS residual / \muK} &
\multicolumn{3}{c}{Nominal RMS residual / \muK} \\ \cline{2-4} \cline{5-7}
Code & $I$ & $Q$ & $U$ & $I$ & $Q$ & $U$ \\
\hline
Polar             & $4.07\times 10^{-3}$ & $3.42\times 10^{-3}$ & $1.22\times 10^{-2}$ & $2.30\times 10^{-3}$ & $2.86\times 10^{-3}$ & $1.27\times 10^{-2}$ \\
MADAM             & $2.09\times 10^{-3}$ & $3.41\times 10^{-3}$ & $1.22\times 10^{-2}$ & $2.00\times 10^{-3}$ & $2.86\times 10^{-3}$ & $1.27\times 10^{-2}$ \\
Springtide        & $2.09\times 10^{-3}$ & $3.41\times 10^{-3}$ & $1.22\times 10^{-2}$ & $2.00\times 10^{-3}$ & $2.86\times 10^{-3}$ & $1.27\times 10^{-2}$ \\
MADmap$^{\dag}$   & $2.16\times 10^{-3}$ & $3.71\times 10^{-3}$ & $1.83\times 10^{-1}$ & $2.02\times 10^{-3}$ & $2.87\times 10^{-3}$ & $1.46\times 10^{-2}$ \\
MapCUMBA$^{\ast}$ & $2.09\times 10^{-3}$ & $3.41\times 10^{-3}$ & $1.22\times 10^{-2}$ & $2.00\times 10^{-3}$ & $2.86\times 10^{-3}$ & $1.27\times 10^{-2}$ \\
MapCUMBA$^{\dag}$ & $1.23\times 10^{-1}$ & $1.70\times 10^{-2}$ & $2.47\times 10^{-2}$ & $5.39\times 10^{-3}$ & $2.88\times 10^{-3}$ & $1.27\times 10^{-2}$ \\
ROMA$^{\ast}$     & $2.09\times 10^{-3}$ & $3.41\times 10^{-3}$ & $1.22\times 10^{-2}$ & $2.00\times 10^{-3}$ & $2.86\times 10^{-3}$ & $1.27\times 10^{-2}$ \\
ROMA$^{\dag}$     & $2.09\times 10^{-3}$ & $3.41\times 10^{-3}$ & $1.22\times 10^{-2}$ & $2.00\times 10^{-3}$ & $2.86\times 10^{-3}$ & $1.27\times 10^{-2}$ \\
\hline
\end{tabular}
\caption{RMS pixel residuals for the polarization simulations without
noise. For the optimal map-making codes ${}^{\ast}$ indicates that
noise correlations were ignored and ${}^{\dag}$ that they were taken
into account; see the text for further details.}
\label{tab:Pno_noise}
\end{center}
\end{table}

\begin{table}[h]
\begin{center}
\begin{tabular}{l c c c c c c}
\hline\hline
     & \multicolumn{3}{c}{Cycloidal RMS residual / \muK} &
\multicolumn{3}{c}{Nominal RMS residual / \muK} \\ \cline{2-4} \cline{5-7}
Code & $I$ & $Q$ & $U$ & $I$ & $Q$ & $U$ \\
\hline
Polar      & $102.91$ & $148.05$ & $148.41$ &  $97.36$ & $138.23$ & $139.31$ \\
MADAM      & $102.86$ & $147.97$ & $148.33$ &  $97.32$ & $138.16$ & $139.25$ \\
Springtide & $106.18$ & $152.51$ & $153.18$ & $100.03$ & $141.91$ & $143.25$ \\
MADmap     & $102.86$ & $147.97$ & $148.34$ &  $97.32$ & $138.16$ & $139.24$ \\
MapCUMBA   & $102.86$ & $147.97$ & $148.33$ &  $97.32$ & $138.16$ & $139.25$ \\
ROMA       & $102.86$ & $147.97$ & $148.33$ &  $97.32$ & $138.16$ & $139.24$ \\
\hline
\end{tabular}
\caption{RMS pixel residuals for the polarization simulations with noise.}
\label{tab:Pnoise}
\end{center}
\end{table}

Polarization map-making introduces the additional complication of
handling pixels in which the $I$, $Q$ and $U$ Stokes parameters are
either mathematically or numerically degenerate. In the white noise
case this degeneracy can be seen in the properties of the $3\times3$
($I,Q,U$) pixel triplet covariance sub-matrices of the full
block-diagonal inverse pixel-pixel noise correlation matrix
$\vec{S}\inv = \vA\tr\vN\inv\vA$. In these runs we used the
determinant criterion of Polar, so that Polar rejected all $(I, Q, U)$
pixel triplets whose covariance matrix had a determinant below some
threshold, while MapCUMBA and MADmap imposed a minimum condition
number on this matrix for the pixel triplet to be accepted.  The
implications of these acceptance criteria, and the choice of threshold
for each, is an ongoing research activity.

\subsection{Expected white noise levels}
\label{sec:whitenoise}

The map residuals in Tables~3 and 5 are due mainly to remaining noise
in the maps.  They can be compared to the expected level of white
noise, which we discuss in this subsection.

The white noise level in our simulated noise timelines was set to
$\sigma = 1279.78~\muK$ for the unpolarized detectors, and $\sigma =
1811.15~\muK$ for the polarized detectors.  A full year (366~days) TOD
from one HFI detector contains $366\times86400\times200 =
6,324,480,000$ samples. If the whole sky were sampled uniformly (same
number of samples from each pixel), we would get $n = 31.4$ hits per
pixel for an $N_\mathrm {side}= 4096$ map, and $n = 125.7$ hits per
pixel for an $N_\mathrm {side}= 2048$ map, from one detector.  This
would lead to a white noise rms level of 228.34~\muK\ (114.17~\muK)
per pixel for an $N_\mathrm {side}= 4096$ ($N_\mathrm {side}= 2048$)
$I$ map from a single unpolarized detector.  Assuming also an optimal
sampling of polarization directions, from the set of four polarized
detectors, the white noise level would be 80.79~\muK\ (114.25~\muK)
per pixel for the $I$ ($Q$, or $U$) map with $N_\mathrm{side} = 2048$.
And from the one-year data from the full set of twelve HFI 217\,GHz
detectors (8 polarized, 4 unpolarized), the white noise level would be
40.38~\muK (80.79~\muK) per pixel for the $I$ ($Q$ or $U$) map with
$N_\mathrm {side}= 2048$.

A standard measure of performance is to give the white noise level for
a pixel whose size corresponds to the resolution of the detector.
From the above numbers the white noise level per a
$5\arcmin\times5\arcmin$ pixel would be 13.87~\muK\ or $\Delta T/T =
5.1\times10^{-6}$ for the $I$ map, and 27.75~\muK\ or $\Delta T/T =
10.2\times10^{-6}$ for the $Q$ and $U$ maps.  For fourteen months of
data these numbers are smaller by a factor of $\sqrt{12/14}$ giving
$\delta T/T = 4.7\times10^{-6}$ for $I$ and $9.4\times10^{-6}$ for $Q$
and $U$, in accordance with the Planck instrument performance goals.

The actual average white noise level will be higher, since the sky
will not be sampled uniformly, and also the polarization directions
will not be sampled optimally for every pixel.  On the other hand,
there will be regions of the sky where the noise will be much lower.
We calculated the resulting theoretical white noise RMS values for our
maps using the actual pointing of the TOD used in this study.

An $N_\mathrm {side}= 2048$ pixel is typically just 1\farcm7 across.
The Planck spin axis is shifted by about 2\farcm5 at every repointing.
The simulations presented here involve either one or four detectors,
rather than the full complement of twelve that Planck will have.  As a
result, the sampling of the sky in these simulations is quite
nonuniform at small scale.  This leaves ``cracks'' between the
scanning rings of a single detector or a 4-detector, 2-horn polarized
pair at low ecliptic latitudes in an $N_\mathrm {side}= 2048$ map.
The expected nutation of the satellite spin axis, included in our
simulated data, reduced this effect somewhat by broadening the
scanning rings, but many pixels were not hit at all, and many more
were hit only a few times.  Pixels hit only a few times, moreover,
were not necessarily hit by all the detectors, and the sampling of
polarization directions may be poor.

For the one-detector case, $189\,146\,022$ of the $201\,326\,592$
pixels of the $N_\mathrm {side}= 4096$ map were hit, with an average
of $n = 33.4$ hits per hit pixel.  If the hits were distributed
uniformly over these pixels, the white noise level would be
$\sigma_\mathrm{map} = 1279.78/\sqrt{n} = 221.32~\muK$.  Actually the
hits are distributed quite nonuniformly, with $1/\langle1/n_p\rangle =
11.4 \ll 33.4 = \langle n_p \rangle$, giving an expected RMS white
noise of $\sigma_\mathrm{map} =
1279.78\times\sqrt{\langle1/n_p\rangle} = 378.95~\muK$ on the map.
Here $n_p$ indicates the number of hits in a pixel, and
$\langle\cdot\rangle$ an average taken over all hit pixels.  We see
that the map residuals in Table~3 are quite close to this expected
level of white noise.

The corresponding results for the four-detector cases with
polarization are presented in Table~6.  Some of the pixels of the
$N_\mathrm {side}= 2048$ map are not hit at all.  In the nominal
scanning strategy there are also large holes at the ecliptic poles,
where the scanning does not extend.  Of the pixels that are hit, we
have defined as \emph{degenerate} those map pixels whose $\kappa_p^{-1}
< 10^{-6}$, and \emph{bad} those with $10^{-6} \leq \kappa_p^{-1} <
10^{-2}$, with the idea that degenerate pixels should be excluded from
the map-making completely, whereas bad pixels can be calculated by the
map-making codes, but are excluded from the calculations of the map
RMS residuals or the expected white noise levels (see Section
\ref{sec:deg_pixs}). Inclusion of bad pixels would increase
dramatically the $Q$ and $U$ residuals, because the low reciprocal
condition number means that noise is amplified in the solution of
these Stokes parameters.

We can see from Table~6 that the main reason the RMS residuals in
Table~5 are significantly larger than in the ideal case of uniform sky
coverage and optimal polarization sampling is the nonuniform sky
coverage.  Nonoptimal sampling of the polarization directions gives an
additional contribution.  This latter contribution is sensitive to the
chosen cut-off (here $10^{-2}$) for ``bad'' pixels.  Using
$\kappa_p^{-1}<10^{-1}$ as a criterion, we would halve the difference
between optimal and actual polarization sampling, at the cost of
leaving out an additional 17196 (5540) pixels in the cycloidal
(nominal) case.

Table~6 cannot be directly compared to Table~5, because the set of
pixels included is not exactly the same.  Different codes used
slightly different criteria for excluding pixels, and Table~5 is
calculated for the common set of pixels calculated by all codes,
which is slightly smaller than the set of $\kappa_p^{-1} \geq
10^{-2}$ pixels.  We have therefore added in Tables~6 and 7 a column
giving the Polar RMS residual for exactly the same set of pixels as
the other entries in Tables~6 and 7. We can see that the expected
white noise level for the actual scanning predicts well this Polar
result, indicating that the RMS residuals for the map-making codes
come mainly from the white noise.
\begin{table}[h]
  \begin{center}
    \begin{tabular}{c c c c c c c}
      \hline\hline
      \multicolumn{2}{c}{ } & & uniform hits & actual hits &
      actual hits & Polar with \\
      \multicolumn{2}{c}{pixels} & & opt.pol.sampling &
      opt.pol.sampling & act.pol.sampling &  $\kappa_p^{-1}\geq 0.01$ \\
      \cline{1-2}
      not hit & discarded & & $\sigma_\mathrm{map} / \muK$ &
      $\sigma_\mathrm{map} / \muK$ & $\sigma_\mathrm{map} / \muK$ &
      $\sigma_\mathrm{map} / \muK$ \\
      \cline{1-7}
      \multicolumn{7}{c}{Cycloidal} \\
      \cline{1-7}
      40383 & 56140 & $I$ &  80.71 & 102.82 & 102.97 & 103.06 \\
      holes 0 & degenerate 54380 & $Q$ & 114.14 & 145.41 & 149.96 &
      150.02 \\
      cracks 40383 & bad 1760 & $U$ & 114.14 & 145.41 & 149.83 & 149.98
      \\
      \cline{1-7}
      \multicolumn{7}{c}{Nominal} \\
      \cline{1-7}
      145626 & 10 346 & $I$ & 80.66 & 97.26 & 97.29 & \\
      holes 144021 & degenerate 7559 & $Q$ & 114.07 & 137.55 & 138.31 &
      \\
      cracks 1605 & bad 2787 & $U$ & 114.07 & 137.55 & 139.43 &  \\
      \hline
    \end{tabular}
    \caption{The expected level of white noise in the maps for the 4-detector
      polarization simulations.
    }
  \end{center}
\end{table}

For the cycloidal scanning strategy we actually had available
simulated TOD for the full set of 12 detectors.  It can be seen from
Table 7 how including 8 or 12 detectors leads to more uniform sky
coverage.  The small-scale nonuniformity due to the cracks between
scanning rings goes away.  The large-scale nonuniformity due to more
visits per pixel near the ecliptic poles remains. The degenerate and
bad pixels, that were due to the cracks have practically
disappeared. Thus the results are much closer to the ideal case.

\begin{table}[h]
  \begin{center}
    \begin{tabular}{c c c c c c c}
      \hline\hline
      \multicolumn{2}{c}{ } & & uniform hits & actual hits &
      actual hits & Polar with \\
      \multicolumn{2}{c}{pixels} & & opt.pol.sampling &
      opt.pol.sampling & act.pol.sampling & $\kappa_p^{-1}\geq 0.01$ \\
     \cline{1-2}
     not hit & discarded & & $\sigma_\mathrm{map} / \muK$ &
     $\sigma_\mathrm{map} / \muK$ & $\sigma_\mathrm{map} / \muK$ &
     $\sigma_\mathrm{map} / \muK$ \\
     \cline{1-7}
     \multicolumn{7}{c}{Cycloidal, 8 polarized detectors} \\
     \cline{1-7}
     0 & 4 & $I$ &  57.12 & 65.72 & 65.72  & 65.79 \\
     holes 0 & degenerate 3 & $Q$ & 80.79 & 92.94 & 93.16 & 93.23 \\
     cracks 0 & bad 1 & $U$ & 80.79 & 97.94 & 93.04 & 93.11 \\
     \cline{1-7}
     \multicolumn{7}{c}{Cycloidal, all 12 detectors} \\
     \cline{1-7}
     0 & 16 & $I$ & 40.38 & 45.88 & 45.88 & \\
     holes 0 & degenerate 3 & $Q$ & 80.79 & 92.94 & 93.16 &
     \\
     cracks 0 & bad 13 & $U$ & 80.79 & 92.94 & 93.04 &  \\
     \hline
    \end{tabular}
    \caption{The expected level of white noise in the maps for
      8-detector and 12-detector simulations.}
  \end{center}
\end{table}

\subsection{Resource requirements}
\label{sec:resource}

The codes described here operate on the full TOD, and their resource
requirements are sizable.  The one-year TOD from four HFI detectors
contained $4\times366\times24\times3600\times200 = 25,297,920,000$
samples.  All codes except Springtide keep the whole dataset in
memory.  Tables~8 and 9 show the memory and CPU times required for
both the one-detector (unpolarized) and four-detector (polarized)
cases. The codes were run on the NERSC \emph{Seaborg} supercomputer,
an IBM SP RS/6000, with clock speed 375~MHz and peak performance
1.5\,Gflops/processor.  Note that for the same code these run times
can vary depending on load and status of the supercomputer.  Since not
all timings were obtained in exactly the same manner, the comparison
should be taken as indicative only.

\begin{table}[h]
  \begin{center}
    \begin{tabular}{l r rr r r}
      \hline\hline
      Code & number of & \multicolumn{2}{c}{memory required} & run
      time & total \\
      & processors & \multicolumn{2}{c}{(reserved) in GB} & (min) & CPU h \\
      \hline
      Polar      & 256 &  60 &  (256) &  23 &   98 \\
      Springtide & 128 &  45 &  (128) &  21 &   45 \\
      MADAM      & 256 &  60 &  (256) &  18 &   77 \\
      MADmap     & 512 & 200 &  (512) &  90 &  768 \\
      MapCUMBA   & 768 & 300 &  (768) &  52 &  666 \\
      ROMA       & 512 & 100 & (1024) & 129 & 1101 \\
      \hline
    \end{tabular}
    \caption{Memory use and run times for the different codes in the
      unpolarized case with one detector. See main text for a discussion.}
  \end{center}
  \label{tab:res_unpol}
\end{table}

\begin{table}[h]
  \begin{center}
    \begin{tabular}{c r rr r r}
      \hline\hline
      Code & number of & \multicolumn{2}{c}{memory required} & run
      time & total \\
      & processors & \multicolumn{2}{c}{(reserved) in GB} & (min) & CPU h \\
      \hline
      Polar           & 512  & 400 & (512)   &  37  &  316 \\
      Springtide      & 512  & 70  & (512)   &  36  &  307 \\
      MADAM           & 512  & 400 & (512)   &  45  &  384 \\
      MADmap          & 2048 & 800 & (2048)  &  90  & 3078 \\
      MapCUMBA        & 1024 & 800 & (1024)  & 100  & 1707 \\
      ROMA            & 1024 & 800 & (2048)  & 206  & 3515 \\
      \hline
    \end{tabular}
    \caption{Memory use and run times for the different codes in the
      polarized case with four detectors.  See text for discussion.}
  \end{center}
  \label{tab:res_pol}
\end{table}

All algorithms solve a linear system by iterative methods.  In
destriping (Polar, Springtide) and destriping-based (MADAM) methods,
the quantities to be solved are a set of baselines for the TOD.  As a
result, the system is smaller than in optimal map-making (MADmap,
MapCUMBA, ROMA). Thus the destriping codes are significantly faster
than the optimal codes.

The processors in \emph{Seaborg} are organized into nodes of 16
processors.  Processors of the same node share memory.  Most nodes
have 16~GB of memory, but some have 32~GB or 64~GB.  Because of the
size of the TOD, the codes require lots of memory.  Since the
communication overhead between processors increases when the same task
is performed with a larger number of processors, run time increases as
memory per processor decreases.  As a result, the 32~GB and 64~GB
nodes are faster.  However, queue times on the high-memory nodes are
longer, so most of the codes were run on the 16~GB nodes.  The ROMA
code requires the 32~GB nodes, because at one point in the calculation
it keeps a full sky map in the memory requested by a single processor.
For the other codes, the memory requirement is dominated by the size
of the TOD, and the variation between codes reflects differences in
how the TOD are stored in memory (single or double precision; how the
pointing and orientation of the detector is represented).  For the
unpolarized case, MapCUMBA was run with an older version of the code,
which required more than twice the memory amount of the current code
and used a slower FFT library.

Springtide requires less memory than the other codes, because it keeps
only ring-maps in memory simultaneously.  A ring-map for a Planck
one-hour pointing period contains many fewer pixels than the number of
samples in the corresponding one-hour TOD.  Springtide could achieve
lower noise in the maps by using baselines shorter than one hour.
This would reduce the compression factor from TOD to ring-maps; for
1-minute baselines, as used by Polar, there would be no compression.
But this would also increase the size of the matrix equation to solve,
increasing computation time.

\section{Conclusions}

The main goal of the simulations reported in this paper was to compare
various map-making codes, and to demonstrate that they can deal with
Planck-size data sets.  The complexity of the simulated data was kept
to a minimum in order to isolate software-induced systematic effects.
No instrumental systematic effects other than noise correlations were
included in the simulated data.  Comparisons based on more realistic
simulations will be made in future papers, which will assess the
impact of strong gradients in the signal due to foregrounds, and
include a more realistic treatment of the instrumental transfer
function, e.g. through the inclusion of beam asymmetries.
Nevertheless, some useful results can be identified.

\subsection{Scanning Strategy}

The Planck design allows considerable flexibility in the choice of the
scanning strategy, even in orbit.  The refinement of the scanning
strategy will therefore be an important aspect of the pre-launch
simulation and analysis work.  Our ability to assess the pros and cons
of various candidate strategies will improve as our map-making
algorithms include the functionality to deal with increasing levels of
realism in the simulated data.

Keeping these qualifications in mind, we compare the map-making
residuals for the nominal and cycloidal scanning strategies. The
results are summarized in Table~\ref{tab:Pnoise}. There is a slight
preference for the nominal strategy. This preference is a result of
the smoother distribution of integration time on the sky for the case
of the nominal scanning.  Comparison of Figures 2 and 3, on the other
hand, shows that the power in the residuals is higher at the lowest
multipoles for the nominal scanning, at least for this particular
realization of the noise TOD.  An accurate quantification of the
effects of scanning strategy requires, however, ensemble averaging,
e.g., using Monte Carlo simulations.  This will be the subject of
future work.  The issue is important for proper understanding of the
mission's ability to measure, for instance, the reionization bump at
low multipole.

Furthermore, we anticipate that the nominal scanning will perform more
poorly than the cycloidal scanning when our simulations contain a more
realistic set of instrumental systematics.  Once these are included,
other features of the scanning strategy will become important, such as
the ability to revisit pixels on a range of time scales in order to be
able to reject systematic effects (such as residuals from
quasi-periodic signals such as cooler noise) and the ability to cross
through pixels in several different directions in order to allow the
reconstruction of the beam transfer functions.

We will return to the assessment of the relative benefits of scanning
strategies in future publications.

\subsection{Resource requirements}

The map-making codes described in this paper require considerable
resources for Planck-sized data.  All codes except Springtide keep the
entire data stream in memory.  Memory requirements are therefore
dominated by the size of the TOD.  Springtide requires
significantly less memory, since it first calculates ring-maps, which
compresses the data by a factor of 20--30 (for a Planck-type scanning
strategy).

The optimal codes (MADmap, MapCUMBA, ROMA) achieve slightly lower
noise in the final maps than Polar and Springtide, but require an
order-of-magnitude more CPU time.  Polar achieved lower noise than
Springtide, because it worked with shorter baselines (1~minute instead
of 1~hour).  Using shorter baselines would increase the memory
requirement of Springtide.  MADAM, which combines optimal map-making
ideas with destriping, achieves practically the same noise levels as
the optimal codes, with similar memory requirements, but in a time
comparable to the destriping codes.

\subsection{Future improvements}

The simulated data prepared for this work are the most advanced so
far, but are still far from the reality of the Planck experiment.
This holds for instrumental systematics as well as for accurate
modeling of the sky signal.  No foregrounds have been included, and
several aspects of the CMB emission have been simplified.  In
particular, no gravitational lensing or tensor signals have been
included.  Both processes have their main impact on the B modes of
polarization anisotropy.  Lensing is a well-understood and inevitable
effect in cosmology, distorting the CMB and producing B modes in a
broad peak in the power spectrum centered at a $\ell\approx1000$, with
an amplitude much smaller than that of E modes.  Tensor signals show up
primarily as degree-scale B modes.  Early reionization could make a
tail of that component appear on very large angular scales, with an
amplitude which might be detectable by Planck.  The pattern of the
total intensity anisotropies on scales of about three degrees or more
has been taken directly by the WMAP data.  This affects also a
component of the E polarization mode pattern.  The angular extension
and reliability of the WMAP pattern in our simulation may
certainly benefit from the future releases of WMAP data.

\begin{acknowledgements}
  The authors would like to thank the University of Helsinki for its
  hospitality in June 2004 when the CTP Working Group met to undertake
  this work.
  MAJA is a member of the Cambridge Planck Analysis Centre, supported
  by PPARC grant PPA/G/R/1997/00837.
  Part of this work was supported by NASA.  CB was
  partly supported by the NASA LTSA Grant NNG04CG90G.
  This research used resources of the National Energy Research
  Scientific Computing Center, which is supported by the Office of
  Science of the U.S. Department of Energy under Contract
  No. DE-AC03-76SF00098.
  We acknowledge the use of version 0.1 of the Planck reference sky model,
  prepared by the members of Planck Working Group 2 and available at
  {\tt http://www.planck.fr/heading79.html.}
  EK and TP were supported by the Academy of Finland grants no.
  75065, 205800, 213984, and 214598. TP wishes to thank the
  V\"{a}is\"{a}l\"{a} Foundation for financial support.
  Some of the results in this paper have been derived using the
  HEALPix~(G\'{o}rski et al.~\cite{gorski05}) package.
\end{acknowledgements}

\end{document}